# Magnetic properties of CrSnS$_3$: A new Van der Waals ferromagnet

F. Obad and Kh. A. Ziq[*]

*Physics Department, College of Engineering and Physics*

*Interdisciplinary Research Center for Advanced Materials, King Fahd University of Petroleum & Minerals, Dhahran 31261, Saudi Arabia*

## Abstract

We report an experimental discovery of CrSnS$_3$, a new Van der Waals ferromagnetic (FM) with Curie temperature $T_C$ ~119K. The Curie temperature is in qualitative agreement of Chittari et. al. DFT prediction ($T_C$ =112.3K) for an Ising model with the addition of Coulomb potential. The FM ordering temperature in CrSnS$_3$ is the highest among Cr-based Van der Waals materials: CrI$_3$, CrSiTe$_3$ and CrGeTe$_3$. The dc-susceptibility revealed a sharp increase at Tc (~119K) indicative of a first order transition; then raises to a wide maximum at AFM Neel temperature $T_N$~90K. The ac-susceptibility reveals the presence of two sharp peaks; one close to Tc (119K) and the other at $T_N$~90K. Ac-susceptibility measurements in an applied dc-magnetic field indicates that CrSnS$_3$ undergoes two successive phase transitions: a sharp increase in the susceptibility indicates a first order PM to FM at 119K, followed by more gradual decrease at the onset of AFM transition at $T_N$~90K. The magnetization isotherms below 30K indicates that the materials orders in an AFM state; while the magnetic isotherms at T >30K rises sharply to its saturated values confirming the ferromagnetic state.

[*]Corresponding author: kaziq@kfupm.edu.sa

1- Introduction

The discovery of the graphene renewed research interest in magnetic properties of various 2-dimensional van der Waals materials and their properties. The coupling between the layers is week; while the in-plane coupling is mainly strong covalent bonding. In particular; ferromagnetic properties of these materials are being intensively investigated for their possible use in spintronics and magnetic semiconductors [1, 2, 3]. Various properties of these materials such as electrical and magnetic are directly influenced by these strong and weak coupling.

A widely investigated classes of materials are the Cr-based compounds; namely: Cr$_2$Ge$_2$Te$_6$, CrSbSe$_3$ and CrI$_3$ [2]. The hope that this may lead to a better understanding of ferromagnetism in one and two dimensions and that may facilitate their usage in spintronics and other technological application[4-10].

The transition-metal trichalcogenides commonly have ferromagnetic insulating ground state and share the general chemical formula MTX$_3$ where M is magnetic transition element (V, Cr, Mn, Fe, Co or Ni) and X is Chalcogens (S, Se, Te) [13-19]. There are two main groups in this family of materials. Group A materials (MAX$_3$) with A is element in group IV (Si, Ge, Sn); and group B (MBX$_3$) with B an element in group V (P, As, Sb) [11,12].

The general structure of the MAX$_3$ compounds is stacked layers as ABC and held together mainly by van der Waals forces [12]. The bilayer structure of the compound consists of M$^{3+}$ cations and (A$_2$X$_6$)$^{-6}$ anions. The crystal structure varies depending on how these ions are arranged [11]. The A atoms have less electrons compared to B atoms, yielding compounds with more negativity and larger nominal metal cation valences 3$^+$ rather than 2$^+$ in the MBX$_3$ compounds. The anions (B$_2$X$_6$)$^{-4}$ of group V are replaced by (A$_2$X$_6$)$^{-6}$ of group IV. This may lead to more in-plane negativity hence stronger 2-dimensional behavior and stronger anisotropy effects. Chittari et. al. used DFT calculations for an Ising model and predicted that CrSnS$_3$ is metallic ferromagnet with $T_C$ =70.7 K [21]. Moreover, Chittari et. al. found that the addition of Coulomb potential switches the metallic state into half-metals, and it leads to semiconducting and enhances FM ordering temperature to $T_C$ =112.3K [21]. Similar enhancement of $T_C$ has been predicted for CrSnSe$_3$ and CoSnS$_3$.

The MAX$_3$ structures is expected to have quasi two-dimensional layered crystal structure rather than a pseudo one-dimensional crystal structure, with edge-sharing, double rutile chains in MBX$_3$ compounds [12, 22, 23]. The quasi two-dimensional layered crystal structure shows distinctive phase diagram in an applied field. As follows from the Mermin-Wagner theorem, the 2D isotropic Heisenberg magnet red no ordered state at any finite temperature above $T_C$ = 0 K, while for the 2D system with XY anisotropy, the topological Kosterlitz and Thouless ordering may take place at a critical temperature $T_C \neq 0$ K [24, 25].

In this work, we report an experimental discovery of CrSnS$_3$ Van der Waals ferromagnetic material with relatively high $T_C$ ~ 119K which is almost double the Curie temperature found other Cr-based FM van der Waals material such as CrI$_3$ or CrSbSe$_3$. This material has a 2D unit cell composed of two Cr$^{3+}$ ions and one [Sn$_2$S$_6$]$^{6-}$ [14, 26].

2- Experimental

High purity elements (99.99%) of Cr, Sn and S are mixed in 1:1:3 molar ratio have been used to prepare CrSnS$_3$ using solid state reaction. The elements were mixed, grinded pressed in pellets and sealed under partial Argon pressure in quartz tube. To avoid exploding the quartz tube, it is necessary to initially anneal the pellets at a temperature lower than the boiling point of sulfur (445 ºC), then use slow heating rate (~10-20 ºC /hour) to higher temperatures. We used 400 ºC and 600 ºC as intermediate temperatures. The pellets were annealed at 400 ºC for 24 hours, then slowly heated to 600ºC and annealed for 48 hours then cooled to room temperature. The pellets were re-grinded, pressed in pellets then reannealed at ~1200 ºC for 24 hours then furnace cooled to room temperature.

X-ray diffraction patterns were obtained using Broker D2 diffractometer with Cu-Kα radiation (λ=1.5405Å). Magnetic ac-susceptibility ($\chi_{ac}$) was measured using a homemade ac-susceptometer incorporating a SR830 DSP Lock-In Amplifier and LakeShore temperature controller. The excitation field is $H_{ac}$ ~ 0.33 Oe and the oscillating frequency $f$ = 887 Hz. In addition, the ac-susceptometer hosts a set of modified design of Maxwell coils that provides highly uniform DC-field over the sample space. The magnetic isotherms were performed using a computer-controlled 9-Tesla PAR 4500/150A variable temperature vibrating sample magnetometer (VSM).

3- Results and Discussion

The variations of the magnetization ($H_{dc}$=200Oe) with temperature is presented in Fig. 1. The figure shows a rapid rise at Curie temperature $T_C$ ~119K, reaching a wide maximum near Neel temperature $T_N$~90K, then drops with a shoulder near 15K. The sharp raise indicates a first order ferromagnetic transition at $T_C$~119 K while the maximum at $T_N$~90K indicates an antiferromagnetic ordering.

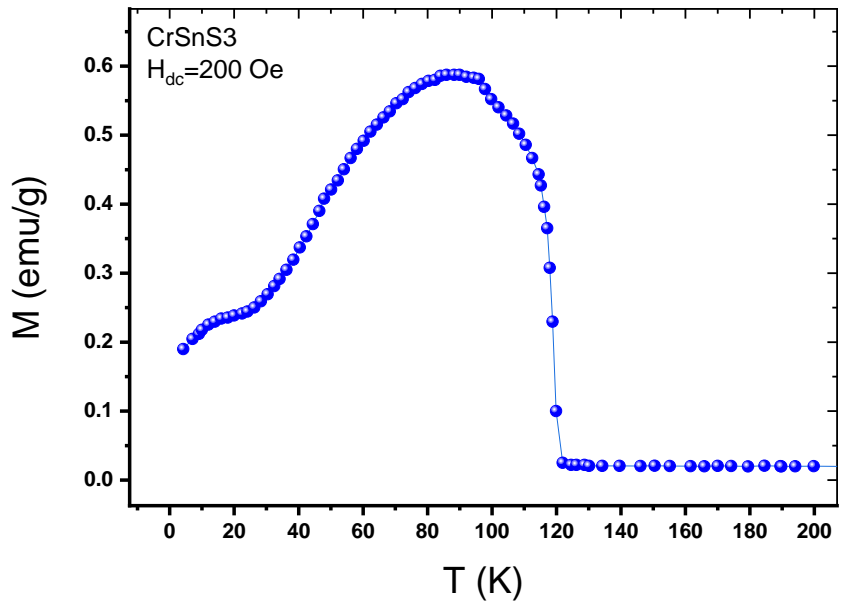

Fig. 1. Variation of the magnetization (at H=200Oe) versus temperature.

The temperature dependence of the real and imaginary parts of ac-susceptibility ($\chi_{ac}$) -measured at $f$=887Hz and Hac~0.33 Oe- for CrSnS$_3$ are shown in Fig. 2a and the derivative of real part ($\chi'$) in Fig. 2b. The sharp increase in the ac-susceptibility at $T_C$ ~119K indicates the onset of FM transition. The derivative of the susceptibility (Fig. 2b) also reveals a very sharp ferromagnetic transition at $T_C$=119K and a FWHM ~ 0.5K. The sudden rise in the ac-susceptibility to a sharp peak clearly indicates a first order phase transition. The ac-susceptibility reveals another small peak near $T_N$~90K, the antiferromagnetic ordering temperature; then drops to a shoulder like near 25K.

As the temperature decreases below $T_N$, the susceptibility decreases rapidly then raises to clear maximum near 90K (Fig. 2a.).

This peculiar behavior is also reflected in the derivative of the susceptibility (Fig. 2b.) which reveals another maximum at 90K indicating yet another transition to possibly antiferromagnetic state. Below 90K, the susceptibility continues to drop becoming almost constant at 25K. Similar behavior can also be seen in the imaginary part of the susceptibility ($\chi''$). As shown in Fig. 2a.

The magnetic state can be further investigated by looking at the effect of dc-magnetic field bias on the ac-susceptibility. The results are presented in Fig. 3. The applied dc-field gradually suppresses the sharp peak and shifts it to lower temperatures. As the dc-field increase, the ac-susceptibility shows a wide maximum in the temperature range (~85-110K), a behavior commonly seen in ferromagnetic dc-susceptibility. The susceptibility drops fast below 80K reaching minimum near 25K. The drop may be caused strengthening of the antiferromagnetic exchange interaction. As the applied dc-field increases, the width of the plateau increases (Fig. 3.), indicating that the materials becomes less responsive to the low $H_{ac}$ (~ 0.33 Oe) and the ac susceptibility in large dc-fields (~40 Oe) looks much similar to the dc-susceptibility (see Fig. 1) [23, 26].

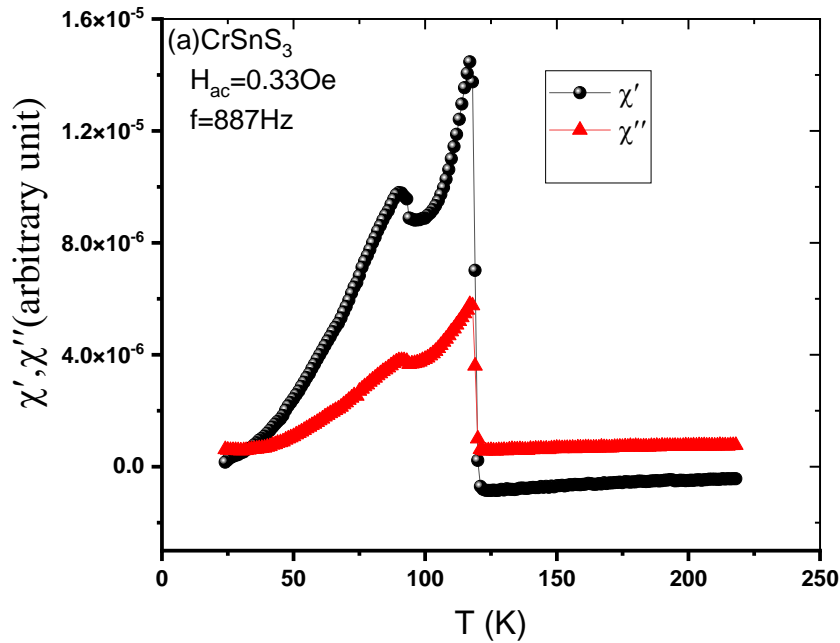

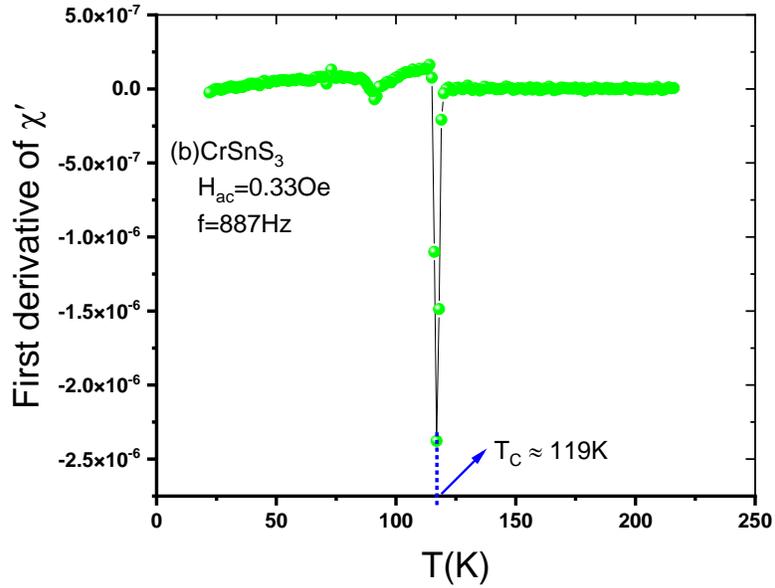

Fig. 2. (a) Variations of the real part of the ac-susceptibility $\chi'$ (circles), and the imaginary part $\chi''$ (triangles). Curves are shifted for clarity of the graph. (b) Temperature variations of the derivative of the real part $d\chi'/dT$.

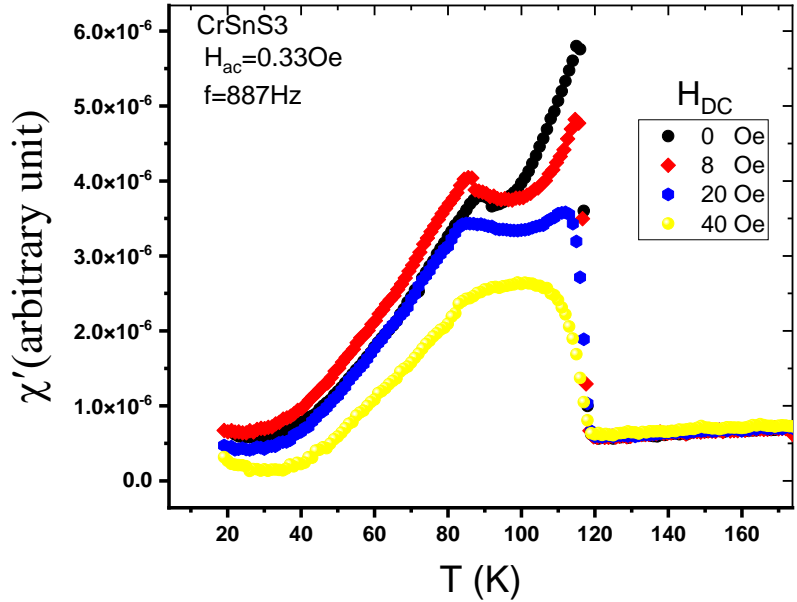

Fig. 3. Temperature variation of in-phase $\chi'$ of ac susceptibility for ($CrSnS_3$) at different dc-magnetic field at fixed $H_{ac}$ = 0.33 Oe and f = 887 Hz.

To further illustrate the effects of the cycling dc-field on the transition temperature we measure the ac susceptibility at a fixed temperature (near $T_C$) while varying dc-field cycled between ±45Oe. The results are presented in Fig. 4. The figure reveals a maximum in the susceptibility at $H_{dc}=0$ that the signal completely disappeared at 120K, while there is a small peak at T=119K. This suggests that the onset of the transition is in between (119-120K).

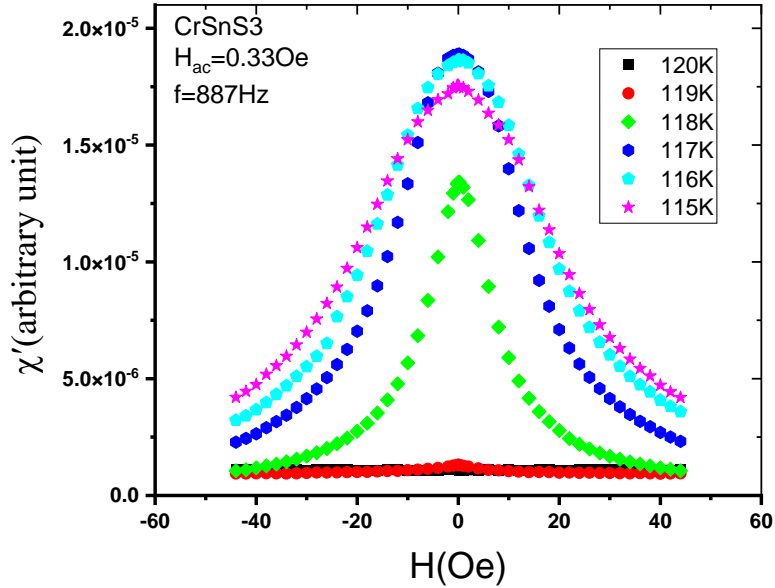

Fig. 4. Variations of in-phase ($\chi'$) of ac susceptibility for CrSnS3 versus dc bias fields at different temperatures.

The isothermal magnetization (*M vs. H*) is presented in Fig. 5. The hysteresis losses are negligible (not shown). At the lowest temperature, (T=4.2K), the magnetization raises gradually then increases linearly, with slight reduction in the slope showing no sign of saturation. This indicates an antiferromagnetic behavior at low temperature. As the temperature increases (~50K), the initial magnetization rapidly increases reaching saturation near 1kOe, indicative a ferromagnetic ordering. This behavior continues up to just below T≈120K. At T >120K; the magnetic state is paramagnetic and the magnetization increases linearly with applied field. The transition from a ferromagnetic to the paramagnetic occurs close to 120K which is in line with the result obtained for ac- and dc-susceptibilities (see Figs. 1 and 2).

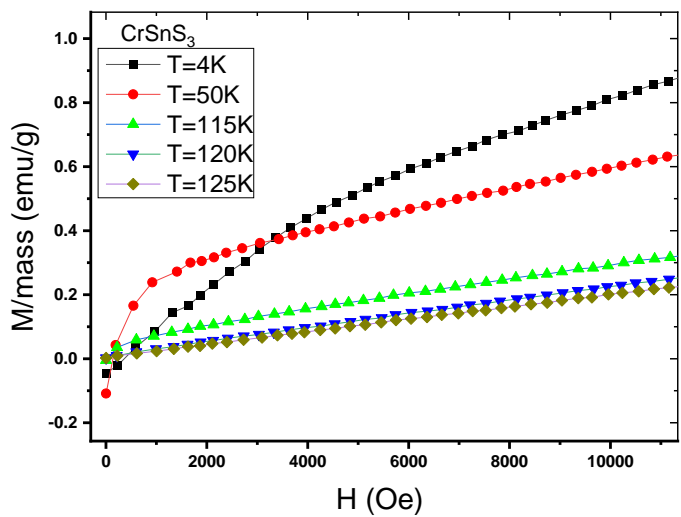

Fig. 5. Variations of the isothermal magnetization with the applied magnetic field (*H*)

### 4- Conclusions

We reported an experimental discovery of a new Cr-based Van der Waals ferromagnetic materials CrSnS$_3$ with a Curie transition $T_C \approx 119K$, which is about double the Curie temperature found in other known Cr-based Van der Waals FM materials. Magnetic isotherms and ac-susceptibility measured in various dc-magnetic field confirmed the FM state in CrSnS$_3$. The FM ordering temperature is in qualitative agreement with the Chittari *et. al.* DFT prediction for an Ising model with the addition of Coulomb potential U (Tc=112.3K). The applied dc-field gradually suppresses the sharp peak and shifts it to lower temperatures. As the applied dc-field increases, the width of the FM transition increases, and the ac-Susceptibility shows a wide maximum in the temperature range (~85-110K). The drop in the susceptibility below 80K indicates strengthening of antiferromagnetic exchange interaction.

**Acknowledgment:** We would like to acknowledge the support of this project provided by King Fahd University of Petroleum & Minerals, Deanship of Scientific Research under project number SB201024.

**Competing Interests:** all authors declare no competing interests.